# Thermoelectric Properties of GaN with Carrier Concentration Modulation: An Experimental and Theoretical Investigation

Ashish Kumar[1,*], Saurabh Singh[2], Ashutosh Patel[3], K. Asokan[1], D. Kanjilal[1].

[1]*Inter-University Accelerator Centre, New Delhi, 110067, India*
[2]*Toyota Technological Institute, Hisakata 2-12-1, Tempaku, Nagoya 468-8511, Japan.*
[3]*Department of Mechanical Engineering, IISc Bangalore, 560012, India*

*email address: ashish@iuac.res.in, dr.akmr@gmail.com

**Abstract:** Present work investigates the less explored thermoelectric properties of *n*-type GaN semiconductor by combined both the experimental and computational tools. Seebeck coefficients of epitaxial thin films of GaN were experimentally measured in the wide temperature range from 77 K to 650 K in steps of ~10 K covering both low and high-temperature regimes as a function of carrier concentration $2 \times 10^{16}$, $2 \times 10^{17}$, $4 \times 10^{17}$ and $8 \times 10^{17}$ cm$^{-3}$. The measured Seebeck coefficient at room temperature was found to be highest, −374 μV/K, at the lowest concentration of $4 \times 10^{16}$ cm$^{-3}$ and decreases in magnitude monotonically (−327.6 μV/K, −295 μV/K, −246 μV/K for $2 \times 10^{17}$, $4 \times 10^{17}$, $8 \times 10^{17}$ cm$^{-3}$, respectively) as the carrier concentration of samples increases. Seebeck coefficient remains negative in the entire temperature range under study indicate that electrons are dominant carriers. To understand the temperature-dependent behaviour, we have also carried out the electronic structure, and transport coefficients calculations by using Tran-Blaha modified Becke-Johnson (TB-mBJ) potential, and semiclassical Boltzmann transport theory implemented in WIEN2k and BoltzTraP code, respectively. The experimentally observed carrier concentrations were used in the calculations . The estimated results obtained under constant relaxation time approximations provide a very good agreement between theoretical and experimental data of Seebeck coefficients in the temperature range of 260 to 625 K.

**Introduction**:

The exponential growth witnessed by GaN-based optoelectronic and power device applications can be attributed to its excellent combination of semiconductor properties (wide bandgap, high breakdown voltage, high electron saturation velocity, etc.) [1-5]. It has been continuously being explored for further new applications, like a substrate for 2D heterostructures, biosensors, and thermoelectric generators (TEGs) [6-8]. Thermal energy in forms of waste heat or metabolic heat is a promising and dependable power source for electronic devices; for example, TEGs are widely being investigated for their ability to convert thermal energy into electricity through Seebeck effect. These materials, therefore, have enormous potential for meeting a part of our energy needs and in enabling self-powered sensors, but improved materials and optimized device geometries are required before this can be realized [9,10]. The thermoelectric (TE) performance of a material can be considered by the parameter, *figure-of-merit* (ZT), defined as $ZT = \sigma S^2 T / \kappa$. Here, σ and κ are electrical and thermal conductivities, $S$ is Seebeck coefficient, and T is the absolute temperature, respectively. Thus, large magnitude of $S$ and $\sigma$ along with low $\kappa$ is an essential condition for better efficiency. Presently, $Bi_2Te_3$ and SiGe based materials are mostly being used in TE devices for commercial and spacecraft applications. Despite the well-known performance, $Bi_2Te_3$ is scarce, toxic and has a low operational temperature (150°C) [8,11]. Similarly, for SiGe, efficiency is low and further improvement is limited due to narrow bandgap which necessitates the research for a better high-temperature TE materials [12]. At high temperatures, the intrinsic minority carriers get excited across the bandgap, shifting the Fermi level towards midgap and resulting in the decreased magnitude of the Seebeck coefficient. A wide bandgap materials tuned with sutable carrier concentrations can prevent such excitation of minority carriers at high temperature and preserve the *n*- and *p*- character of the semiconductor.

For GaN and its clan of alloy, inherent high $S$ values, presence of *n*- and *p*-type doping, high temperature and chemical stability, and nontoxicity provide an advantage over other semiconductor materials [13,14]. The wide bandgap and semiconducting electronic structure allow creating the impurity states in addition to the change in carrier concentrations, which effectively helps in modifying the transport properties of the GaN for its novel industrial applications [12,15,16]. However, one prominent limitation of GaN as first choice thermoelctric materials is that it has high thermal conductivity, which prevents commercial development of such application.

Various approaches like nanostructure, alloying, defect engineering have been used to address these issues making nitrides based on early transition metals and group-13 elements highly relevant[17-19]. An on-chip Peltier cooler device fabricated from GaN in the proximity of a high power transistor might provide heat dissipation solution without adding extra or new material and therefore, preventing contamination in the semiconductor processing line. Such promising applications necessitate detailed investigation and optimization of TE properties. Brandt *et al.* experimentally found large Seebeck coefficients for both *p*- (800 µV/K) in Mg-doped GaN and *n*-type (−350 µV/K) in undoped GaN epitaxial layers grown on sapphire [20]. Based on the electron transport model, Liu and Baladin theoretically predicted the TE properties of GaN and $Al_XGa_{1-X}N$ alloys by considering energy-dependent scattering effect due to all possible sources [21]. The calculated TE parameters pave a direction for experimental optimization such as synthesis conditions for significant improvement in TE properties of GaN-based material by changing the carrier concentrations. Several groups have tried to tune the TE properties of bulk GaN materials by modification of electronic structure either by substitutions of additional elements such as In or Al at Ga site or by lowering the dimensions in the form of thin films or nanowires [22,23]. Both *p*- and *n*-type materials were successfully achieved depending upon the amount or type of the additional impurity introduced in the pure GaN material [24]. Our group previously reported enhancements in the Seebeck coefficient and power factor of *n*-GaN thin films and other materials by optimized ion beam irradiations[25,26]. Yamamoto *et al.* prepared bulk GaN material with micron size of particles, which showed a negative *S* value of −50 µV/K at room temperature that further increased with temperature [27]. In the same investigation, with substitution of In at Ga site in GaN, they observed a low magnitude of Seebeck coefficients while having significant improvement in electrical conductivity and enhancement of overall power factor ($S^2\sigma$). The comparative study on both bulk and thin-film GaN, Kucukgok *et al.* reported a positive and large Seebeck coefficients which found to be decreased with increase in carrier concentrations[28]. However, their study was limited to 300 K. Yamaguchi *et al.* practically demonstrated the free-standing prototype TE device fabricated using GaN, with an open-circuit voltage ($V_{op}$) of 28 mV with power output ($P_{max}$) 3.35 µW at 153 K temperature difference across the device [27]. Sztein *et al.* designed a more efficient device made from Si-doped GaN thin film which had a 0.3 V open-circuit voltage with a maximum output power of 2.1 µW at 30 K temperature difference [14]. An effort is made to tune the Seebeck coefficients by varying the nitrogen content in Si-doped

$GaN_XAs_{1-X}$ thin-film system, and no improvement as compared to GaAs system was noticed even though the density of states effective mass found to be increased with nitrogen concentration, and the possible reason for the degrade in the thermoelectric property was attributed to the scattering effect [29]. By using the Monte Carlo simulation, Davoody *et al.* predicted that *ZT* of GaN material can be achieved to 0.8 at room temperature with optimal doping in 3 nm nanowire[30]. However, experimentally, there are several challenges in the fabrication of the nanowire with such dimensions and tuning the doping amount, and further, implement it into the device level at a relatively cheaper price is yet to be done. Using the first principle calculations, Reshak *et al.* show that GaN can be a suitable candidate for thermoelectric applications as it possesses a high Seebeck coefficient and has the chance of optimizations [29]. By having the look at the previously reported work on the GaN-based material from the thermoelectric applications point of view, we realized that a systematic investigation on thermopower behaviour of this system in the practical operating temperture region of electronic devices, and covering the wide range of carrier concentrations are still lacking to the best of our knowledge. One of the main reasons might be the challenge in the measurement of Seebeck coefficient in insulating materials without proper ohmic contacts. This gives us the motivation to carry out a detailed study on the thermopower performance of GaN materials by taking the samples of a wide range of carrier concentrations. The present findings will be very useful not only in finding the GaN material with appropriate carrier concentration for the suitable TE application but also to provide a basic understanding of the experimentally observed behaviour using the first-principle calculation. This will also help to find new TE materials with wide bandgap semiconductor systems by tuning the carrier concentrations.

In this manuscript, we present a detailed investigation of thermoelectric properties of GaN in a wide temperature range of 77 – 650 K using both experimental and computational tools. Seebeck coefficients measurements on the four different samples with carrier concentrations range ($10^{16}$–$10^{18}$cm$^{-3}$) were carried out. Electronic structure and transport coefficients calculations were also performed to understand the temperature-dependent behaviour. All the samples show *n*-type character and have a large magnitude of Seebeck coefficients. The carrier concentrations, which obtained from the hall set-up measurements, were taken into consideration for the calculations of Seebeck coefficients. Theoretically estimated Seebeck coefficients were found to be in very good agreement with the experimental data in 260 to 625 K temperature range.

**Experimental and Computational Details:**

GaN epitaxial layers grown by metal-organic chemical vapour deposition (MOCVD) technique on the sapphire substrate were used for experimental measurements. The thickness of the GaN layers was approx. 3 μm. The Phillips X'pert PRO diffractometer with Cu-$K\alpha$ source ($\lambda$=1.5418 Å) was used for structural characterization. For electrical and Seebeck coefficient measurements, special ohmic contact pads on samples were fabricated as thermoelectric measurements in insulating or wide bandgap semiconductors (GaN: 3.4 eV) with moderate or non-degenerate concentration are very tricky and so less data is reported. For such materials, electrical measurements cannot be done without proper ohmic contact. Arc contacts or simple metal pressure probes form a non-ohmic junction with high contact resistance and junction potentials which may prevent the accurate measurement of small Seebeck voltages. Very thin ohmic contact pads were deposited in a vacuum ($10^{-7}$ torr) by an e-beam evaporation system at two opposite edges of 10 mm long samples (for Seebeck measurements). We used the Ti/Al/Ni/Au (20/100/20/100 nm) scheme followed by rapid thermal annealing at 1000 K for 30 sec in the $N_2$ atmosphere. Literature shows that the four-layer structure (Ti/Al/Ni/Au) exhibits the lowest quoted contact resistance for *n*-type GaN [2,31,32]. These samples were then mounted in the sample holder for Seebeck coefficient measurements (Figure 1b). Pressure probes containing thermocouples tip were positioned from the top on ohmic contact pads. Temperature-dependent Seebeck coefficient measurement and resistivity measurements were performed using our in-house developed measurement system [33]. For carrier concentration measurements, square-shaped samples with ohmic contacts in four corners were characterised in van der Pauw configuration in an Ecopia system (model HMS 5000).

To understand the thermoelectric properties of GaN, the systematic calculations were performed by use of the first principle method based on the density functional theory. For the calculations of thermoelectric parameters, the semi-classical Boltzmann transport theory based BoltzTraP code is used, where transport coefficients as a function of temperature and chemical potential can be estimated within the constant relaxation time approximations [34]. For the electronic structure calculations, we used the theoretical tools based on the linearised full potential augmented plane-wave (LAPW) + local orbitals (lo) method, implemented into WIEN2k code [35]. Both generalised gradient approximation (GGA) given by PBE, and TB-mBJ exchange-correlation functional were used for the comparison of the energy bandgap and ground state electronic structures [36,37]. For the wz-GaN structure, the Wyckoff positions of Ga atom are (2/3, 1/3, ½) and (1/3, 2/3, 0), whereas N atom occupies the (2/3, 1/3, ½+$u$) and (1/3, 2/3, $u$) positions and the value of $u$ represents the interlayer distance between Ga plane and its nearest N plane in terms of c parameter[38,39]. Experimental lattice parameters were used for all the calculations. To get accurate results of transport properties, the highly-dense k-points mesh within the Brillouin zone (BZ) were set to the 15×15×10, 30×30×25, and 50×50×45 for the optimization, electronic structure, and transport coefficients, respectively. The energy convergence criteria were set to the $10^{-6}$ Rydberg. The muffin-tin radius for the Ga and N was automatically set to the 1.99 and 1.71 Bohr radius, respectively.

**Results and Discussions:**

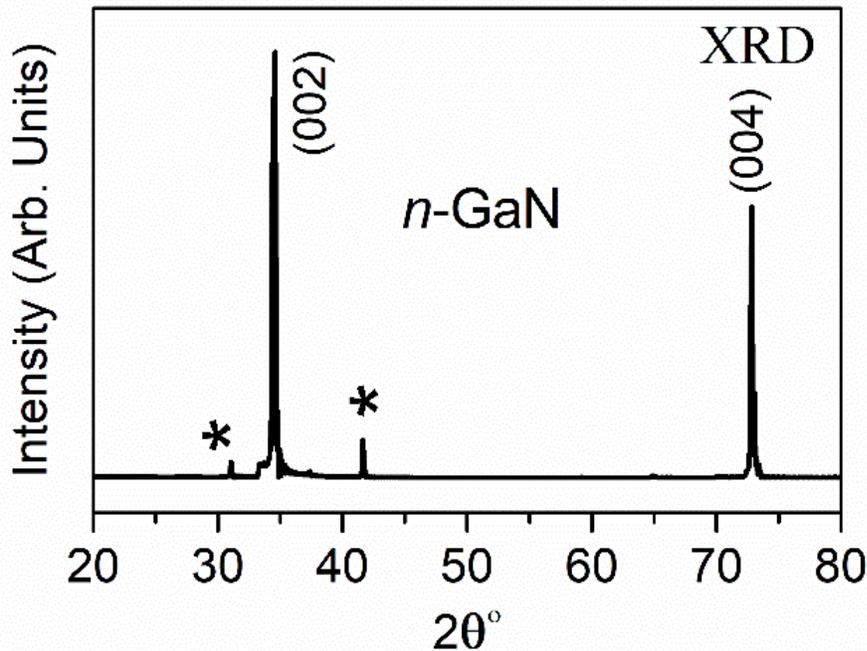

**Figure 1:** XRD data of GaN thin films, showing wurtzite structure orientation (002) along with sapphire substrate peaks (* marked).

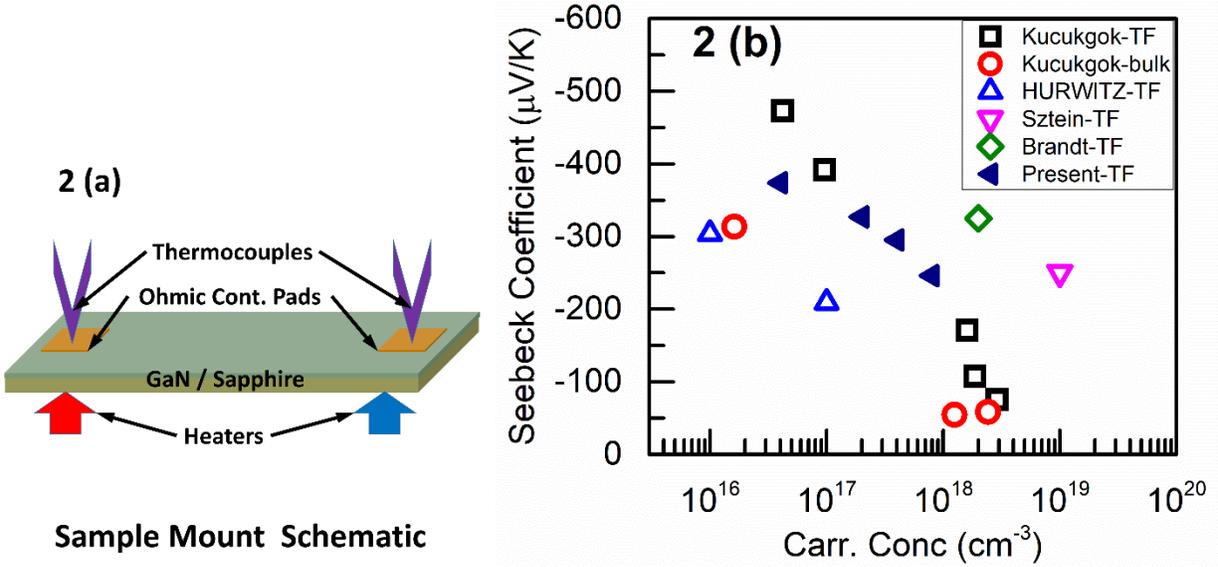

**Figure 2**: (a) Schematic for Seebeck measurement sample having two ohmic contact pads (Ti/Al/Ni/Au) on GaN layer for voltage measurement, (b) Seebeck coefficients measured at 300 K in this study compared with literature of thin films (TF) and bulk GaN [14,15,20,28].

Figure 1 shows the typical XRD curve for epitaxial GaN samples used in present study. The resultant data was matched with standard cell parameter values in JCPDS card -02-1078. The films showed highly oriented (002) peaks due to the wurtzite crystal structure (wz) for GaN. The experimentally measured values of lattice constants '$a$' and '$c$' were found as 3.18 Å and 5.18 Å, respectively. The Seebeck coefficient measurement principle and design are although explained more detailed way and reported in our previous work, it is important to have a brief description of sample characterization[33]. The GaN thin film sample is mounted on keeping contact pads on top (figure 2 (a)). Two copper probes are placed on ohmic contact and holder cup is closed. After creating a low vacuum, the dipstick cryostat is merged in the liquid nitrogen (LN2) dewar. Measurement is done from 77 to 650 K by maintaining a 5 K temperature gradient across two ends of the sample. This system ensures steady-state equilibrium conditions for each data recording. The successive data points for average temperature (mean of hot and cold end temperature of the sample) have a constant difference of approx. 10 K. Instrument calibration is performed by using the constantan standard for the whole temperature range before measuring the GaN samples. The temperature variation of measured Seebeck coefficients of all samples is plotted in the figure 2(b). The negative signs for all samples in full temperature range indicate electrons as majority carriers which is also confirmed by negative Hall coefficients in Hall/resistivity measurements. The observed room temperature Seebeck coefficient value is highest (−374 µV/K) for the lowest carrier concentration sample (4 × 10$^{16}$ cm$^{-3}$) and decreases with increase in doping concentration (−327.6 µV/K, − 295 µV/K, − 246 µV/K for 2 × 10$^{17}$, 4 × 10$^{17}$, 8 × 10$^{17}$ cm$^{-3}$, respectively) which is in accordance the literature cited [28,40,41]. The reported reasons for such behaviour is ascribed to increased scattering factor due to carriers and dislocation density defects. Also, the magnitude of S increases with increasing temperature for all samples. Similar characteristics are observed by various research groups [20,25]. The detailed theoretical explanation for observed data is explained in the later sections.

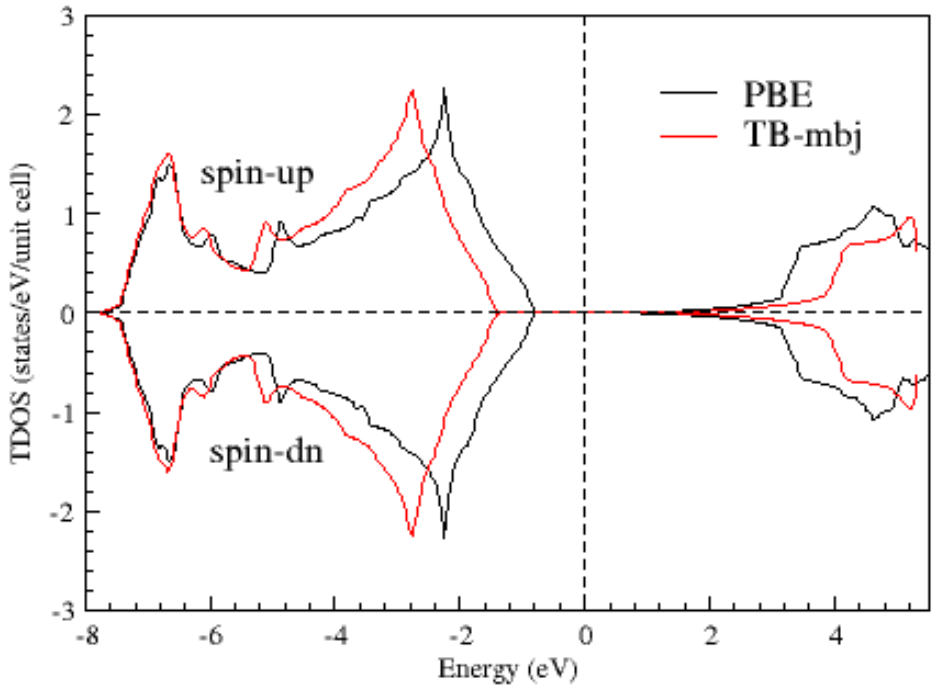

**Figure 3:** The total density of states plot for wurtzite GaN calculated using PBE and TB-mBJ exchange-correlation functional.

Figure 3 shows the total density of states (TDOS) plot obtained from the two different exchange-correlation functional using PBE and TB-mBJ. The calculated bandgap is found to be ~1.7 eV and 2.9 eV for PBE (GGA) and TB-mBJ, respectively. In comparison to the experimental band gap (3.4 eV), the value of estimated bandgap is about half for the case of PBE functional, while in case of TB-mBJ, we could be able to obtain much better agreement [42,43]. The underestimation of bandgap using LDA/GGA functional is a very common issue as reported earlier for the several semiconductor materials. To overcome this problem, there could be two possible solutions by which the value of the bandgap can be improved. First, to impose the correlation effect of on-site Coulomb interaction ($U$) on the $d$-electrons and tune the values of $U$ such that the calculated bandgap becomes equivalent to the experimental value. The second way to improve the computed bandgap is to do the calculation beyond the LDA and GGA approximations by using hybrid functionals such as PBE0, HSE03, HSE06, GW, BZW-EF, etc. [42]. In both cases one can get improvement in the band gap value, however, for the primary case the need of matching bandgap by adopting the different parameter of $U$ value without its experimental validation is a type of fitting, and later case demand high computational facility and also very time consuming, which limits us to use it in the present study. Here, it is important to note that the transport properties are mainly dependent on the electronic structure near the band edge. Once the accurate electronic structure is obtained, the slightly underestimated bandgap can also be corrected which provided the results of transport coefficients in good agreement with experimental results. Therefore, accurate electronic structure is more important than the exact matching of computational bandgap on the cost of time and computational facility. The additional correction in the potential term given by TB-mBJ provides better electronic structure information as compared to the LDA and GGA, and ground-state self-consistency calculation using this exchange functional followed by Boltzmann transport calculation explains the thermoelectric properties in several thermoelectric materials. So our main focus here is to obtain the bandgap in better agreement with the experimentally measured value of bandgap and study the TE properties. The further discussions related to the electronic structure and transport properties will be associated with the results obtained by using the TB-mBJ functional.

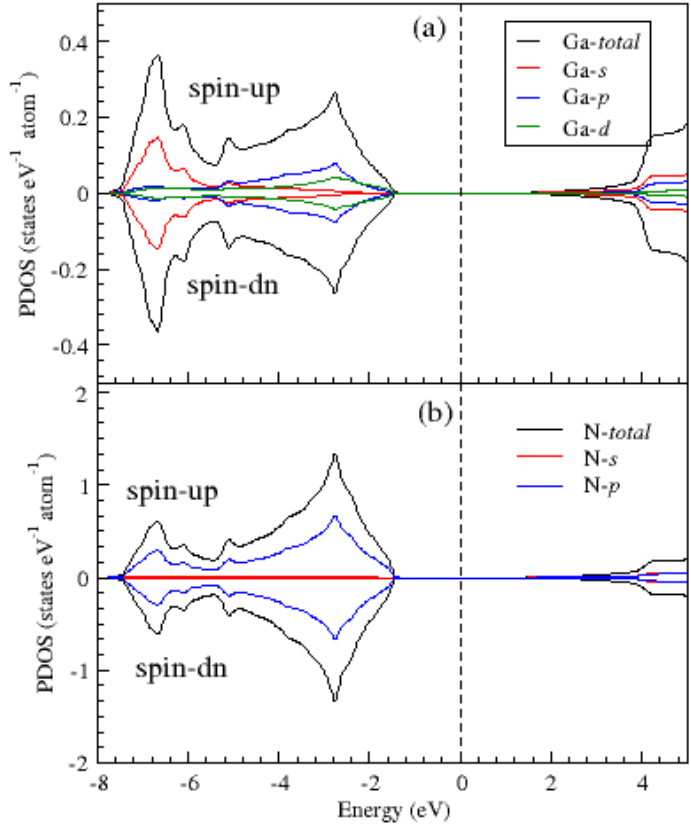

**Figure 4:** The partial density of states (PDOS) (a) for Ga and (b) for N calculated using TB-mBJ functional.

From figure 3, the contribution in TDOS is observed in a wide energy range between −1.6 eV to −7.2 eV for VB. Here, it is to be noted that, the Fermi level ($\mu = 0.0$ eV) is set to the middle of the gap as ground state electronic structure properties are calculated without consideration of the temperature effect. TDOS increases sharply at valence band edge towards lower energy range and the maximum of TDOS value is ~2 states/eV/unit cell at the energy of ~0.8 eV below to the top of VB edge. Two main regions which contributed to TDOS in VB are −1.6 eV to −4.6 eV and −5.8 eV to −7.6 eV, among this two-region the energy range close to the top of the valence band will mainly contribute into the transport properties. The finite density of states starting from the conduction band edge to higher energy range up to 4eV is noticed. This suggests that thermally excited charge carriers from the states near to the top of the valence band can get occupied in the available states at the bottom of the conduction band. To know the contributions in the total density of states from the different orbital, we have also estimated the partial density of states for Ga and N atoms as shown in figure 4a and 4b, respectively. From PDOS shown in figure 4a, we can see that in the VB, the total contributions from the Ga atom for the region −1.6 eV to −4eV, is mainly from the Ga-*p* and Ga-*d* orbitals, whereas for the PDOS of N atom it is mainly from the N-*p* orbital. For the region far away from the VB edge, i.e. −5.6 eV to −7.6 eV, contributions in PDOS for the Ga atom is mainly from the Ga-*s* orbital, in the same region contributions in PDOS for N atoms are also from the N-*p* orbital. From the PDOS, it is evident that strong hybridization occurs between Ga-*p* and N-*p* states in the region −1.6 to −3.0 eV region, while in the region of −5.6 eV to −7.6 eV the hybridization is in between Ga-*s* and N-*p* states. Magnusan *et al.* also have previously reported similar hybridization using X-ray spectroscopy and first-principle methods[44]. By looking at PDOS contribution, we can clearly understand that, in the thermoelectric properties, the main contributions will be from the Ga-*p*, Ga-*d* and N-*p* states as they have major contributions in the density of states near the top of VB edge. The other orbitals Ga-*s* and N-*s* near valence band edges have very small contributions in the PDOS, thus contributions from these orbitals in the transport properties are expected to be very small [45].

To understand the thermoelectric properties of semiconductor materials, information extracted from band dispersion play a key role. Therefore, we have calculated the band dispersion along with the high symmetry points **A** (0, 0, 1/2), **L**(1/2, 0, ½) **M** (1/2, 0, 0), **ϒ** (0 0 0), and **H** (1/3, 1/3, ½), as shown in figure 5 [46]. Both the top of valence band and bottom of the conduction band are at the ϒ point, which describes the direct bandgap semiconductor characteristics of the GaN. The values of the direct bandgap from the dispersion curve, 2.9 eV, is well agreed with that obtained from the TDOS calculation. In comparison to the experimental band gap, ~3.4 eV, and several reports based on DFT calculations the bandgap obtained in our case is in good agreement [42,43]. In the VB, three bands **B1**, **B2**, and **B3** meet each other at the ϒ point suggests that top of the valence band is triply degenerate. Along ϒ-**M** direction, degeneracy lifted and these bands separated from each other at **M** points and become completely non-degenerate, while along ϒ-**A** direction **B2** and **B3** remain merged to each other and meet again with band **B1** at **A** points shows that triplet degeneracy remains same as that of ϒ points. The *k*-point, other than ϒ, at which energy states are closest to the top of the VB is the point **A,** where three bands contribute with the same energy and can be next most probable energy states in the contributions to transport behaviour as it is ~0.4eV below to the VB maxima. At the ϒ point, the bottom of the conduction band has a single band suggesting that the conduction band is non-degenerate. The second next band from the bottom of CB is at ϒ point, and about ~2.5 eV higher energy. Therefore, from the band dispersion, one can expect that the contributions in transport properties will be mainly from the three degenerate valence bands (**B1**, **B2** and **B3**) and one non-degenerate conduction band (**B4**).

For the qualitative understanding of TE properties, effective mass plays an important role, as within the free-electron theory approximation, Seebeck coefficients have the direct dependence on the effective mass of charge carriers describe as, $S = (8\pi^2 k_b^2/3eh^2)m^*T(\pi/3n)^{2/3}$, where $k_b$, $e$, $h$, $m^*$, $T$ and $n$ are Boltzmann constant, electronic charge, planks constant, effective mass, absolute temperature, and carrier density, respectively [8]. This expression suggests that for a given temperature and carrier density, charge carriers having higher effective mass will mainly dominate and results in large Seebeck coefficients. Thus, we have calculated the effective masses at ϒ point along with several different symmetric directions for the three valence bands (**B1**, **B2**, **B3**) and the single conduction band (**B4**). For the estimations of $m^*$, we did the fitting of the equation, $1/m^* = d^2E(k)/h^2 dk^2$, using the symmetric E vs $k$ parabolic region data points centred at ϒ point and along with **A, L, M** and **H** directions[47,48]. The estimated effective masses for holes and electrons are given in table 1. Effective masses for the holes in VB found to be large as compared to the electrons effective masses in the conduction band. This suggests that, for the intrinsic GaN materials, hoes are the dominant charge carriers, and thus, large and positive Seebeck coefficients are expected. In our case, calculated effective masses for both holes and electrons are in very good agreement with the previously reported values using the experimental and theoretical calculations[43,49-51]. As the temperature increases, the holes from all the three degenerate band get thermally excited and get accommodated in the available states in the non-degenerate conduction band. The number of thermally excited electrons from the top of the valence band to the bottom of the conduction band is proportional to the term exp[−$E_g/2K_bT$][52]. At finite temperature, the electrons get thermally excited from the valence band to create holes in the valence band and reach to the conduction band contribute in the thermo emf. As the bandgap of the intrinsic GaN material is larger than the conventional thermoelectric materials. Thus, it is always challenging to populate enough number of thermally excited electrons for significantly large contributions in transport properties. Despite large bandgap, having the semiconducting nature and highly degenerate valence band provides a better chance to improve the transport behaviour by doping with external elements or changing the carrier concentrations of GaN materials by creating defects with tuning appropriate synthesis conditions. In the present investigation, purposefully, we select the four different samples of electron doping and have a large variation in carrier concentration obtained at different synthesis conditions.

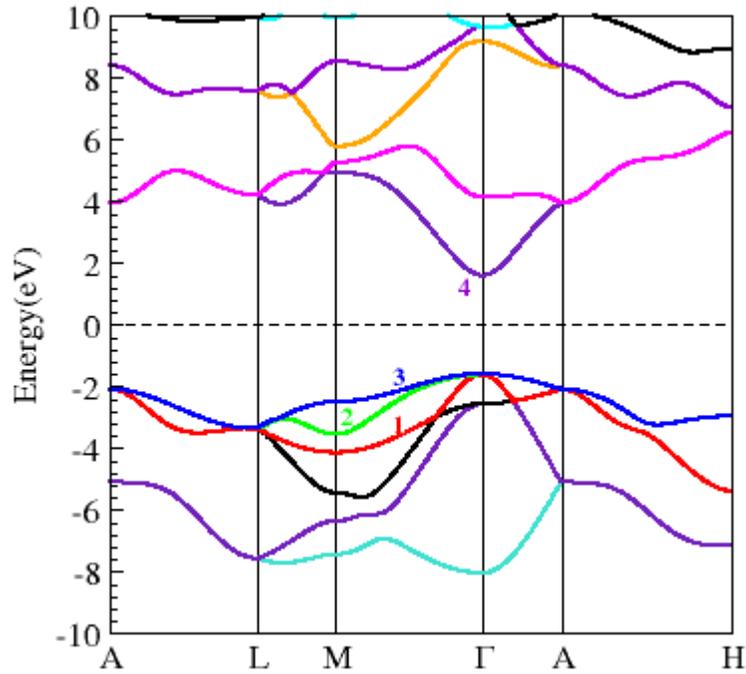

**Figure 5.** Dispersion curve along with the high-symmetry points **A-L-M-ϒ-A-H**

| High symmetric points | Valance bands | | | Conduction band |
|---|---|---|---|---|
| | **B1** | **B2** | **B3** | **B4** |
| ϒ$_{ϒA}$ | 0.54 | 1.84 | 1.84 | 0.24 |
| ϒ$_{ϒM}$ | 0.58 | 1.74 | 1.76 | 0.18 |
| ϒ$_{ϒL}$ | 0.55 | 1.63 | 1.68 | 0.22 |
| ϒ$_{ϒH}$ | 0.52 | 1.62 | 1.70 | 0.23 |

**Table 1**: Effective masses $m^*/m_e$ of holes (for bands **B1, B2 and B3**) and electrons (band **B4**) at ϒ points calculated along the several high symmetric k-points of A, M, L and H.

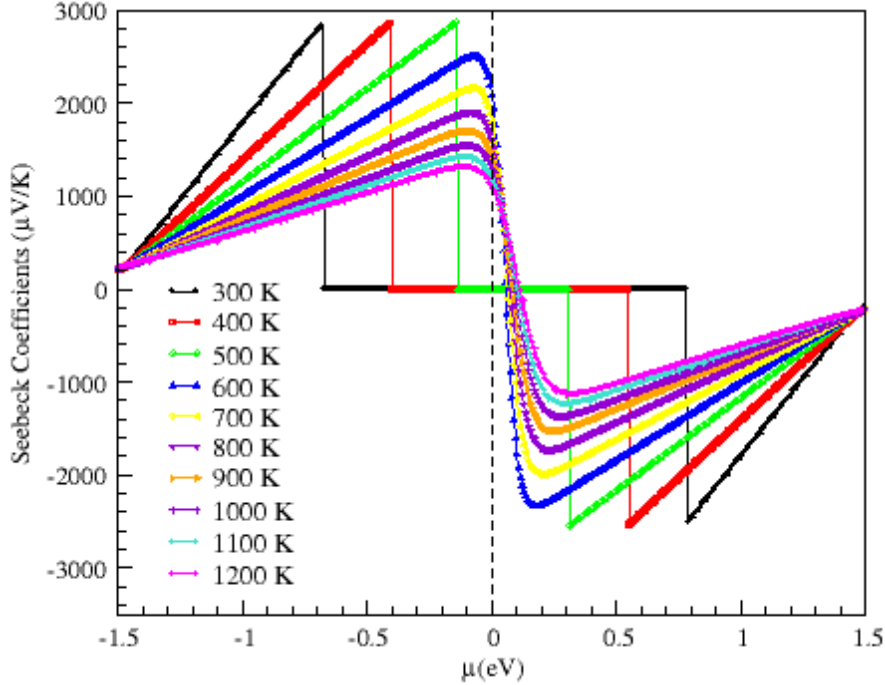

**Figure 6:** The variation of Seebeck coefficients as a function of chemical potential ($\mu$) at different temperatures.

To know the effect of doping on thermoelectric properties, the variation of Seebeck coefficients as a function of chemical potential ($\mu$) at different temperatures in figure 6. It is observed that for the intrinsic GaN, µ=0 eV, the Seebeck coefficients are found to be positive and well consistent with the previously reported experimental and computational data[29]. The step-like behaviour within the bandgap region is found for the plot corresponding to the temperature of 300 K, 400 K and 500 K. Such type of step curve might be due to the unoccupied energy state in the bandgap region, and with such band gap value, energy derivative of the Fermi-Dirac distribution function does not provide the significant contribution in the value of the Seebeck coefficient[29]. As the GaN is found to be thermally stable at high temperature, therefore we computed the Seebeck coefficients up to 1200 K, which can be possible to measure by the commercial experimental setup. For both *p*- ($\mu$<0 eV) and *n*-type (µ>0 eV), the magnitude of Seebeck coefficients are found to be very large (S = 2000 µV/K). Also, we noticed the decrease in the magnitude of Seebeck coefficients occur as temperature increases from 600K to 1200 K. Even at 1200 K temperature, the maximum estimated value show as large as 1000 µV/K. The most important point we noticed are the maxima of Seebeck coefficients corresponding to different temperature plot are within 100 meV from the µ = 0 eV, and the position of µ for the maxima of Seebeck coefficients for each temperature does not change so effectively. This also suggests that tuning the thermoelectric properties for both *p*- and *n*-type is possible for the GaN material.

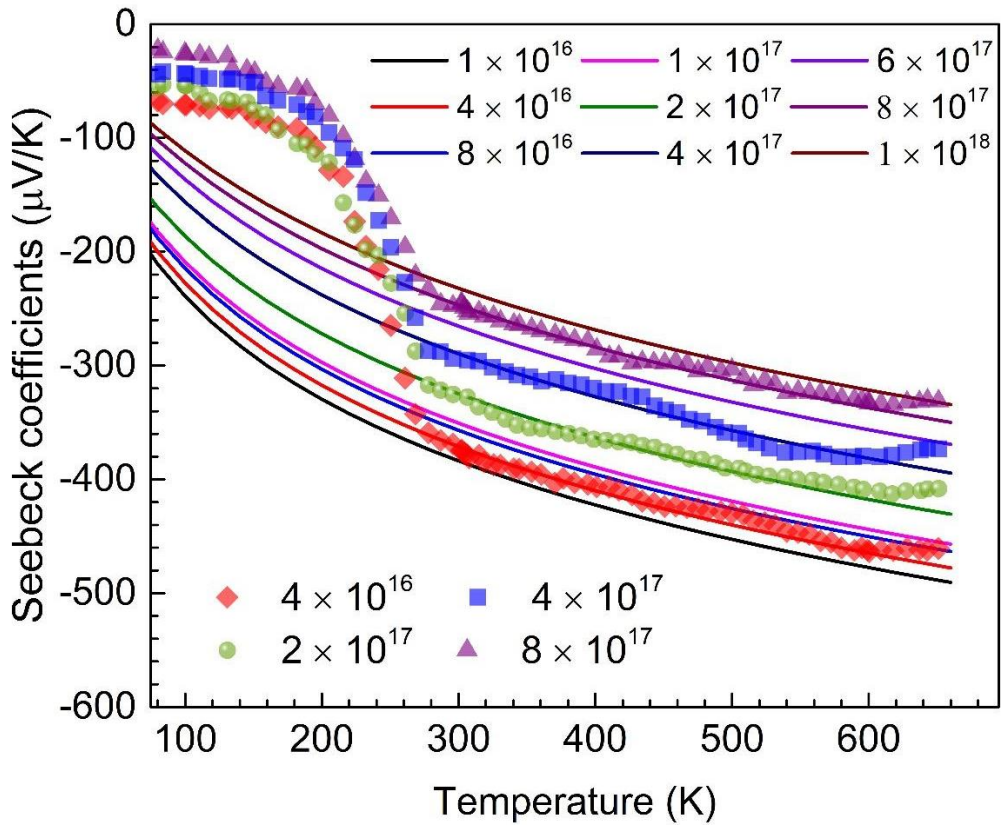

**Figure. 7:** Seebeck coefficient vs temperature. Calculated Seebeck coefficient (solid line) and experimental Seebeck (symbol) for different samples having carrier concentration.

The experimental Seebeck coefficients along with calculated values at different carrier concentrations are shown in figure 7. To understand the experimental data in the wide temperature range, we have calculated the Seebeck coefficients at different carrier concentrations in $1\times10^{16}$ – $1\times10^{18}$ cm$^{-3}$ range. It is seen in figure 7, the experimental and calculated values have a very good match in the 260 – 620 K temperature range. Most interestingly, we observed the good agreement between experimental and estimated Seebeck coefficients, when the values of carrier concentrations used in BoltzTraP calculations are taken same as that of measured values obtained from the Hall set up at room temperature. This clearly shows that consideration of carrier concentration parameters in transport coefficient calculations are very effective and provide better results. In the very high-temperature range, 625–650 K range, we noticed a slight deviation between calculated and experimental data, which can be attributed to the change in carrier concentration with temperature. Experimental trends of almost every sample have an upward curve and have a slight decrease in magnitude in the temperature range 625–650 K. This is quite obvious as Seebeck coefficients are a function of carrier concentrations. With the increase in carrier concentration, the magnitude of the Seebeck coefficient decreases. In the temperature range below 260 K, we noticed a sharp decrease in the magnitude of the Seebeck coefficient with the decrease in the temperature. Also, the scattering factor can play some role in decreasing the magnitude. This is an experimental observation and has also been reported[20]. The large deviation in calculated and experimental data in the low-temperature region is observed. Although we have taken very fine K-mesh 50×50×45 points for accurate electronic structure and transport coefficient calculations, the observation of deviation in low temperature can be due to the limitation of the computational tool. GaN samples are very useful for the applications near room temperature and above, so in this study, we mainly focused our investigation in the high-temperature range. We successfully characterized the TE properties of GaN samples with different carrier concentrations, and understand the temperature-dependent Seebeck coefficient behaviour using electronic structure and transport coefficient calculations.

In our calculations, we have not used any scattering parameter-dependent calculations as per the limitation of the BoltzTraP code. In future, it will be quite interesting to measure the Seebeck coefficient data at much higher temperature range as the GaN sample is thermally and electrically very stable and to understand the transport behaviour with theoretical modelling will be major investigation aspects.

**Conclusions:** In summary, *n*-GaN thin film samples were investigated for fundamentally important thermoelectric properties at a wide temperature range (77 – 650 K). Seebeck coefficient was measured as a function of temperature for four different concentrations below degenerate concentrations for GaN. The observed behaviour complies with the literature. A detailed electronic structure and transport coefficients calculations using codes BoltzTraP and WIEN2k were carried out to understand the experimental data. The estimated temperature dependence of Seebeck coefficients for all samples is found to be in very good agreement with the experimentally observed behaviour for the temperatures (260 – 620 K). This investigation suggests that by tuning the carrier concentrations and optimized thermal conductivity, GaN can be used as potentail *n*-type thermoelectric material for a wide temperature range.

**Acknowledgments**: AK would like to thank Dept. of Science and Technology, India for providing financial support for this work through DST-INSPIRE faculty scheme.

**Data Availability Statement:** The data that support the findings of this study are available from the corresponding author upon reasonable request.

**References:**

[1] S. Kako, C. Santori, K. Hoshino, S. Götzinger, Y. Yamamoto, and Y. Arakawa, Nature materials **5**, 887 (2006).
[2] H. Morkoç, *Handbook of nitride semiconductors and devices* (Wiley-VCH; John Wiley (distributor), Weinheim,, 2008), V: 1-3.
[3] H. Morkoç, *Nitride semiconductors and devices* (Springer, Berlin ; New York, 1999), Springer series in materials science v. 32.
[4] A. Kumar, S. Arafin, M. Amann, and R. Singh, Nanoscale Res Lett **8**, 1, 481 (2013).
[5] A. Kumar, S. Vinayak, and R. Singh, Current Applied Physics **13**, 1137 (2013).
[6] H. Ohta, S. W. Kim, S. Kaneki, A. Yamamoto, and T. Hashizume, Advanced science **5**, 1700696 (2018).
[7] A. S. Yalamarthy, H. So, M. Muñoz Rojo, A. J. Suria, X. Xu, E. Pop, and D. G. Senesky, Advanced Functional Materials **28**, 1705823 (2018).
[8] G. J. Snyder and E. S. Toberer, Nature Materials **7**, 105 (2008).
[9] H. Park, D. Lee, G. Park, S. Park, S. Khan, J. Kim, and W. Kim, Journal of Physics: Energy **1**, 042001 (2019).
[10] D. Narducci, Journal of Physics: Energy **1**, 024001 (2019).
[11] B. Poudel *et al.*, Science **320**, 634 (2008).
[12] N. Lu and I. Ferguson, Semiconductor Science and Technology **28**, 074023 (2013).
[13] H. Ohta, A. Sztein, S. P. DenBaars, and S. Nakamura, (Google Patents, 2014).
[14] A. Sztein, H. Ohta, J. Sonoda, A. Ramu, J. E. Bowers, S. P. DenBaars, and S. Nakamura, Applied Physics Express **2**, 111003 (2009).

[15] E. N. Hurwitz, M. Asghar, A. Melton, B. Kucukgok, L. Su, M. Orocz, M. Jamil, N. Lu, and I. T. Ferguson, J. Electron. Mater. **40**, 513 (2011).
[16] A. Sztein, J. Haberstroh, J. E. Bowers, S. P. DenBaars, and S. Nakamura, Journal of Applied Physics **113**, 183707 (2013).
[17] P. Eklund, S. Kerdsongpanya, and B. Alling, Journal of Materials Chemistry C **4**, 3905 (2016).
[18] B. Biswas and B. Saha, Physical Review Materials **3**, 020301 (2019).
[19] Y. Feng, E. Witkoske, E. S. Bell, Y. Wang, A. Tzempelikos, I. T. Ferguson, and N. Lu, ES Materials & Manufacturing **1**, 13 (2018).
[20] M. Brandt, P. Herbst, H. Angerer, O. Ambacher, and M. Stutzmann, Physical Review B **58**, 7786 (1998).
[21] W. Liu and A. A. Balandin, Journal of Applied Physics **97**, 123705 (2005).
[22] C. Guthy, C.-Y. Nam, and J. E. Fischer, Journal of Applied Physics **103**, 064319 (2008).
[23] H. Tong, H. Zhao, V. A. Handara, J. A. Herbsommer, and N. Tansu, 2009), pp. 721103.
[24] C. Wan, Y. Wang, N. Wang, W. Norimatsu, M. Kusunoki, and K. Koumoto, Science and Technology of Advanced Materials **11**, 044306 (2010).
[25] A. Kumar, J. Dhillon, R. C. Meena, P. Kumar, K. Asokan, R. Singh, and D. Kanjilal, Applied Physics Letters **111**, 222102 (2017).
[26] A. Bhogra *et al.*, Scientific Reports **9**, 14486 (2019).
[27] S. Yamaguchi, R. Izaki, N. Kaiwa, and A. Yamamoto, Applied Physics Letters **86**, 252102 (2005).
[28] B. Kucukgok, B. Wang, A. G. Melton, N. Lu, and I. T. Ferguson, physica status solidi (c) **11**, 894 (2014).
[29] A. H. Reshak, RSC Advances **6**, 72286 (2016).
[30] A. H. Davoody, E. B. Ramayya, L. N. Maurer, and I. Knezevic, Physical Review B **89**, 115313 (2014).
[31] S. Rouvimov *et al.*, Applied Physics Letters **69**, 1556 (1996).
[32] A. Kumar, M. Kumar, R. Kaur, A. G. Joshi, S. Vinayak, and R. Singh, Applied Physics Letters **104**, 133510 (2014).
[33] A. Kumar, A. Patel, S. Singh, A. Kandasami, and D. Kanjilal, Review of Scientific Instruments **90**, 104901 (2019).
[34] G. K. H. Madsen and D. J. Singh, Computer Physics Communications **175**, 67 (2006).
[35] P. Blaha, K. Schwarz, F. Tran, R. Laskowski, G. K. H. Madsen, and L. D. Marks, The Journal of Chemical Physics **152**, 074101 (2020).
[36] J. P. Perdew, K. Burke, and M. Ernzerhof, Physical Review Letters **77**, 3865 (1996).
[37] F. Tran and P. Blaha, Physical Review Letters **102**, 226401 (2009).
[38] D. Andiwijayakusuma, M. Saito, and A. Purqon, Journal of Physics: Conference Series **739**, 012027 (2016).
[39] Y.-N. Xu and W. Y. Ching, Physical Review B **48**, 4335 (1993).
[40] A. Sztein, J. E. Bowers, S. P. DenBaars, and S. Nakamura, Journal of Applied Physics **112**, 083716 (2012).
[41] A. Sztein, H. Ohta, J. E. Bowers, S. P. DenBaars, and S. Nakamura, Journal of Applied Physics **110**, 123709 (2011).
[42] Y. I. Diakité, S. D. Traoré, Y. Malozovsky, B. Khamala, L. Franklin, and D. Bagayoko, arXiv preprint arXiv:1410.0984 (2014).
[43] R. B. Araujo, J. S. d. Almeida, and A. F. d. Silva, Journal of Applied Physics **114**, 183702 (2013).
[44] M. Magnuson, M. Mattesini, C. Höglund, J. Birch, and L. Hultman, Physical Review B **81**, 085125 (2010).
[45] L. Ley, R. A. Pollak, F. R. McFeely, S. P. Kowalczyk, and D. A. Shirley, Physical Review B **9**, 600 (1974).
[46] M. I. Aroyo, A. Kirov, C. Capillas, J. Perez-Mato, and H. Wondratschek, Acta Crystallographica Section A: Foundations of Crystallography **62**, 115 (2006).

[47]	S. Singh and S. K. Pandey, Philosophical Magazine **97**, 451 (2017).
[48]	S. Singh, R. K. Maurya, and S. K. Pandey, Journal of Physics D: Applied Physics **49**, 425601 (2016).
[49]	P. Rinke, M. Winkelnkemper, A. Qteish, D. Bimberg, J. Neugebauer, and M. Scheffler, Physical Review B **77**, 075202 (2008).
[50]	J. S. Im, A. Moritz, F. Steuber, V. Härle, F. Scholz, and A. Hangleiter, Applied Physics Letters **70**, 631 (1997).
[51]	M. Steube, K. Reimann, D. Fröhlich, and S. J. Clarke, Applied Physics Letters **71**, 948 (1997).
[52]	N. Ashcroft and N. Mermin, Solid State Physics **239**, 562 (1976).